# Diffractive flat lens enables extreme depth-of-focus imaging

*Sourangsu Banerji,[1] Monjurul Meem,[1] Apratim Majumder,[1] Berardi Sensale-Rodriguez[1] and Rajesh Menon[1, 2, a)]*

[1]Department of Electrical and Computer Engineering, University of Utah, Salt Lake City, UT 84112, USA.

[2]Oblate Optics, Inc. San Diego CA 92130, USA.

[a)] rmenon@eng.utah.edu

**ABSTRACT**

A lens performs an approximately one-to-one mapping from the object to the image planes. This mapping in the image plane is maintained within a depth of field (or referred to as depth of focus, if the object is at infinity). This necessitates refocusing of the lens when the images are separated by distances larger than the depth of field. Such refocusing mechanisms can increase the cost, complexity and weight of imaging systems. Here, we show that by judicious design of a multi-level diffractive lens (MDL) it is possible to drastically enhance the depth of focus, by over 4 orders of magnitude. Using such a lens, we are able to maintain focus for objects that are separated by as large as ~6m in our experiments. Specifically, when illuminated by collimated light at $\lambda$=0.85mm, the MDL produced a beam that remained in focus from 5mm to ~1500mm from the MDL. The measured full-width at half-maximum of the focused beam varied from 6.6μm (5mm away from MDL) to 524μm (1500mm away from MDL). Since the sidelobes were well suppressed and the main-lobe was close to the diffraction-limit, imaging with a horizontal X vertical field of view of 20º x 15º over the entire focal range was demonstrated. This demonstration opens up a new direction for lens design, where by treating the phase in the focal plane as a free parameter, extreme depth-of-focus imaging becomes possible.

# INTRODUCTION

The lens, an optical element that focuses incident collimated light into a focal spot, is one of the key elements of an imaging system. The distance range measured normal to the lens over which the spot remains tightly focused is generally referred to as the depth of focus (DOF). There is a direct relationship between the DOF and the focal spot size, W, given by: DOF ~ $4W^2/\lambda$ ~ $\lambda/NA^2$, where $\lambda$ is the illumination wavelength and NA is the numerical aperture of the lens. When used for imaging, this results in the images of objects outside the equivalent range in the object side to be blurred. There are obviously many reasons to extend this range for enhanced DOF imaging. This may be achieved via wavefront coding or computationally or via some combination of these [1]. Wavefront coding [2, 3] and related approaches such as with a logarithmic asphere [4] generally leads to a trade-off between resolution and DOF, sometimes at the expense of image quality primarily due to the increased side lobes. The axicon can be used to enhance DOF, but the image resolution and field of view are heavily curtailed [5, 6]. Lenses with multiple discrete foci have also been demonstrated [7, 8]. Optimized apodizers in the pupil plane [9], binary phase optimized phase masks by themselves [10-12] and combined with refractive lenses [13] have been used to enhance DOF. In all cases, the enhancement of DOF is relatively small, less than 1 order of magnitude. Many of these prior examples require extensive post processing; their Field of View (FOV) is heavily curtailed or require multiple elements and tend to be quite thick. Computational enhancement of DOF also suffers from noise amplification and high sensitivity to the signal-to-noise ratio of the collected signals. Table S1 (Supplementary Information) summarizes relevant prior work in this field, from which we can conclude that the best enhancement of DOF over the diffraction limit is ~600 [14], which was achieved with a fractal zone plate combined with a conventional focusing lens. Here, we show that with appropriate design of a flat multi-level diffractive lens (MDL), it is possible to increase the DOF to over $10^4$ times that of the diffraction-limited case as described below. Our MDL is designed to focus at the diffraction limit over the entire range of $z=f_{min}$ to $z=f_{max}$, where z is the distance measured from the MDL (see Fig. 1A). We also note that the NA is defined generally as $\sin(\tan^{-1}(z/R))$, where $z \in \{f_{min}, f_{max}\}$. In other words, the NA of our MDL varies with z as expected from the diffraction limit. The largest NA of our MDL occurs at $z=f_{min}$ and is 0.18, and the DOF enhancement over the diffraction-limited case for this NA is $10^6 \lambda/(\lambda/NA^2) > 3 \times 10^4$, which is orders of magnitude larger than anything demonstrated before. With such a large DOF, it is possible to remove focusing mechanisms from cameras, thereby reducing cost, weight and associated complexity.

**DESIGN**

Our design methodology is inspired by our recent demonstration that for intensity imaging, phase in the focal plane could be treated as a free parameter [15]. In principle, this would enable infinite solutions to the ideal lens problem and one can choose the desired solution based upon other requirements such as achromaticity [15] or enhanced DOF (here). By only constraining the intensity to be focused in a large focal range and allowing the phase within this focal range to vary, we can solve a nonlinear inverse problem via optimization. The details of our design methodology are similar to what has been reported before [16-22]. To summarize, we maximize the focusing efficiency of the MDL by selecting the distribution of heights of its constituent rings (see Fig. 1B). In the case of ExDOF MDL, the focusing efficiency defined as the ratio of power inside a spot of diameter = 3 X FWHM to the incident power, is computed for each focal plane within the desired focal range. The focal range is defined as the range of distances measured from the MDL over which the light is desired to be focused ($f_{max} - f_{min}$ in Fig. 1A). The selection of ring heights is based upon a gradient-descent-directed binary search. We designed, fabricated and characterized an MDL for $\lambda = 0.85\mu m$ with the following parameters: $f_{min} = 5mm$, $f_{max} = 1200mm$, and aperture = 1.8mm. The design was constrained to a ring-width of $3\mu m$, and at most 100 height levels with a maximum height of $2.6\mu m$.

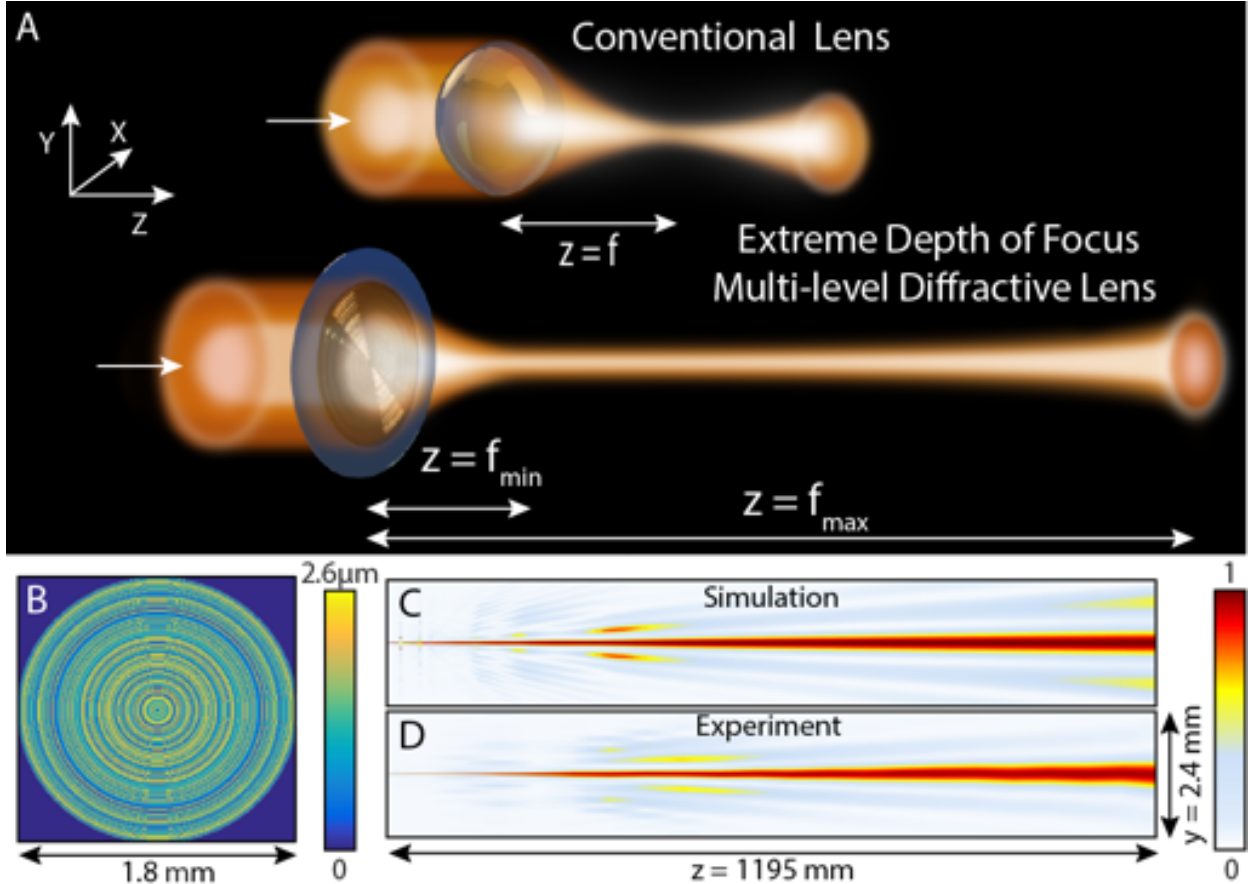

*Figure 1: A. Schematic of a flat multi-level diffractive lens (MDL) that exhibits Extreme Depth-of-Focus (ExDOF) imaging. B Geometry of MDL with focal range = 5mm to 1200mm, aperture = 1.8mm. The C simulated and D measured intensity distributions in the XZ plane for the MDL with focal lengths = 5 mm to 1200mm. The operating wavelength is 850nm*

Note that the f# (NA) varies due to the extended focus between 2.78 (0.18) and 555.56 (0.0009) for the MDL. The diffraction-limited DOF given by $\lambda/\max\{NA\}^2$ is $26\mu m$ for this specific design. Therefore, the enhancement in DOF over the diffraction limit given by $(f_{max}-f_{min})/(\lambda/\max\{NA\}^2)$

is 3.8 X 10$^4$. These enhancements are several orders of magnitude larger than anything that has been demonstrated before and post-processing of the images is not required. We emphasize that here we are ensuring that a single flat MDL is able to span such a large range of f #s and NAs, which, in other words, leads to the extreme depth of focus in imaging. Diffraction limit dictates that for focal spots that are farther away from the MDL will be larger.

**RESULTS AND DISCUSSION**

The light intensity distribution in the XZ plane (see Fig. 1A for coordinate definition) was simulated and plotted for the MDL in Fig. 1C. From the simulation, it is clear that the mainlobe of the focus is maintained over the desired focal range. The full-width at half-maximum (FWHM) of each MDL was seen to be close to diffraction-limited across the focal range. For instance, the FWHM was 524μm at a distance of 1200mm away. In fact, simulations indicate that $f_{max}$ extends to over 1200mm, although sidelobes start appearing (see Fig. 1C). The corresponding measurements are shown in Fig. 1d. In all cases, the sidelobes are also fairly well suppressed within the desired focal range, which is unlike what one would expect from a Bessel beam, for instance. This is extremely important for imaging. The material dispersion of a positive-tone photoresist, S1813 (Microchem) was assumed in the design and simulation [15]. The device was then fabricated using grayscale lithography as has been reported previously (Fig. S6) [15, 17-20]. Note that due to our fabrication constraint, we were limited to a maximum ring height of 2.6μm. Optical micrograph of the fabricated MDL is shown in Fig. 2A.

Each MDL was illuminated by a collimated beam from a supercontinuum source (NKT Photonics SuperK EXB-6) equipped with a tunable filter (NKT Photonics SELECT) that was tuned to 850nm with a bandwidth of 15nm. The point-spread function (PSF or the light intensity distribution in the XY plane) was recorded directly on a monochrome CMOS image sensor (DMM 27UP031-ML, The Imaging Source). The image sensor was placed on a stage and the PSFs were captured at different distances from the MDL (see Supplementary Information for details). After that, these images were concatenated to create the XZ slice that is shown in Fig. 1D. From these distributions, we experimentally confirmed that the incident light remains focused in the designed focal range, 5mm to 1200mm. We further note that the experiments agree with the simulations relatively well.

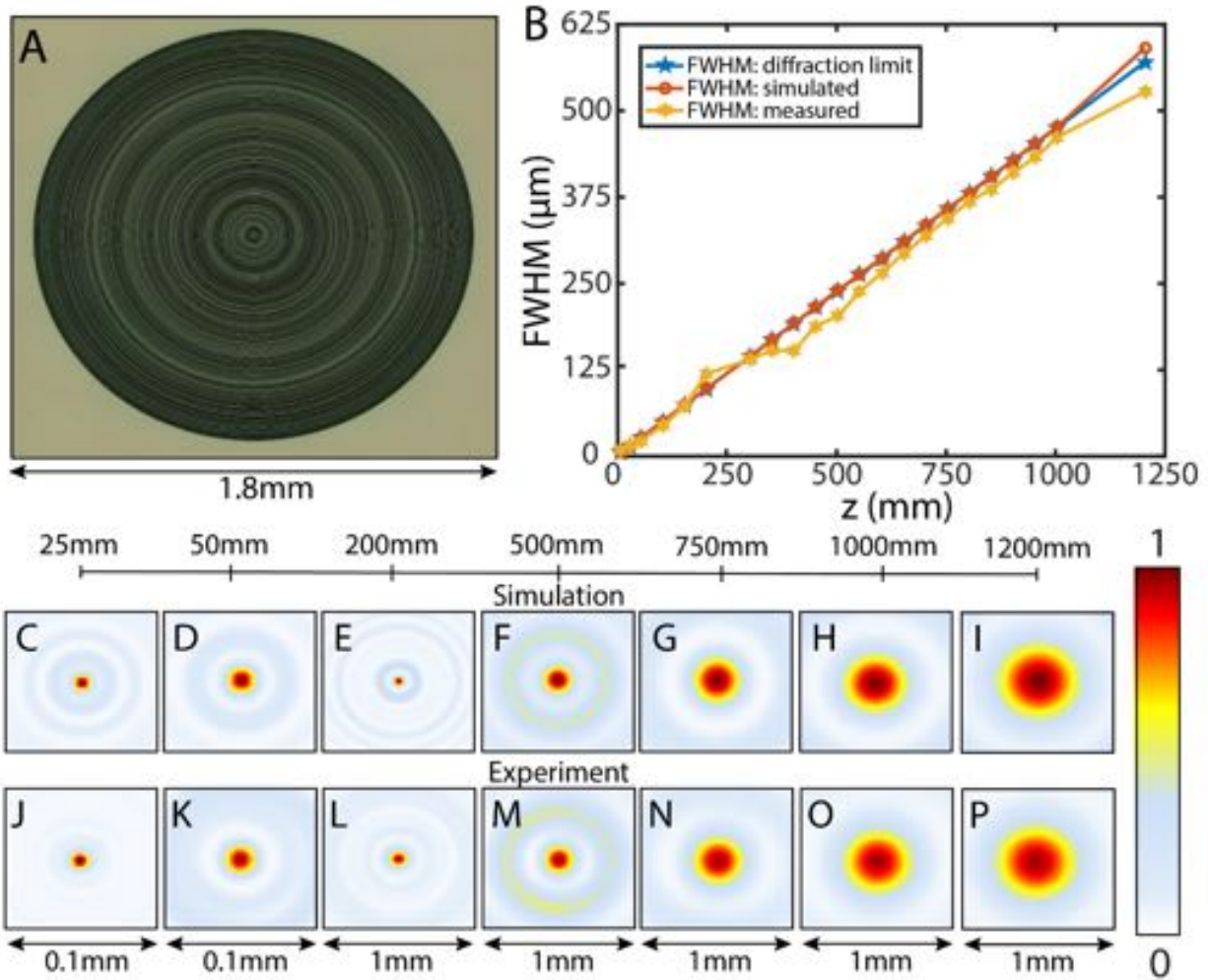

*Figure 2: A. Optical micrographs of fabricated MDL. B. Measured, simulated and diffraction-limited full-width at half-maximum (FWHM) as function of z. C-I Simulated and J-P Measured point-spread functions (light intensity distributions in the XY planes) as function of z, the distance from the MDL.*

The FWHM was extracted from both the simulated and the measured PSFs and plotted as a function of z in Fig. 2B. The diffraction-limited FHWM is also plotted for comparison. We note the excellent agreement between all 3 plots. We summarize the simulated (Figs. 2C-I) and measured (Figs. 2J-P) PSFs for z (distance between the MDLl and the image sensor) = 25mm, 50mm, 200mm, 500mm, 750mm, 1000mm and 1200mm, respectively. The recorded PSFs for some of the other z values are also shown in Fig. S1 [15]. There is good agreement between the experiments and simulations, although we believe some of the discrepancies can be attributed to expected fabrication errors (see section 8 of the Supplementary Information). In any case, it is clear that the MDL focus incident light close to the diffraction limit over its designed focal range. We also computed the modulation transfer function and the focusing efficiency as a function of z from the measured PSFs and summarized the data in Figs. S7 and S2, respectively (Supplementary Information).

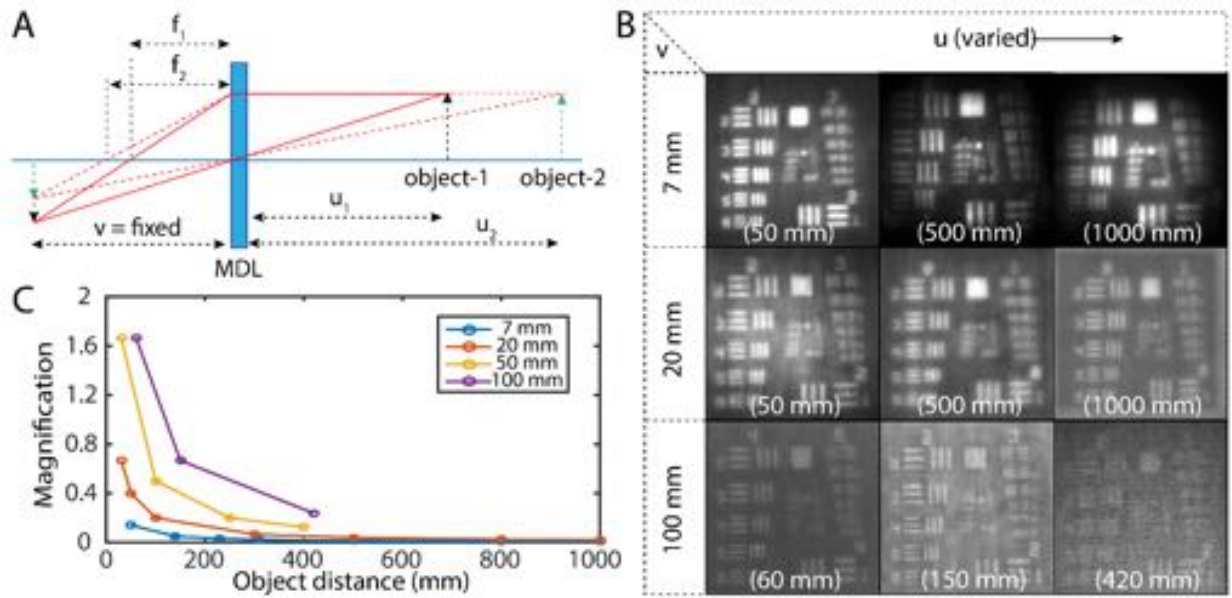

*Figure 3:* *Imaging different object distances without refocusing.* ***A.*** *Focused images of objects at different distances u from the MDL are formed at the same image distance, v.* ***B*** *Images of the AirForce resolution chart for fixed v and varying u. The value of v is fixed for each row and value of u is noted in parenthesis in each image.* ***C*** *Magnification as a function of u for various values of v extracted from the recorded images.*

Finally, we used the MDL to capture test images using the same CMOS image sensor as before. In each case, the exposure time was adjusted to ensure that the frames were not saturated. In addition, a dark frame was recorded and subtracted from the images. In all experiments, we recorded images of the Air Force resolution target in two different scenarios. First, we kept the distance between the MDL and the image sensor (v) fixed and varied the distance between the MDL and the object (u) as illustrated in Fig. 3A. The idea was to demonstrate a camera that does not need to be refocused as u changes. The recorded images for the MDL are shown in Fig. 3B. The values of u are noted in parenthesis in each image. The experiment was repeated for 3 different values of v as indicated. We note that the MDL was able to form focused images for u from 50mm to 1000mm without any change in v, *i.e.*, without having to refocus. From the recorded images, we can compute the magnification of the camera and plot it as function of u and v for the MDL in Fig. 3C. We note that the MDL allows one to change the magnification without any change in the image distance (v). The magnification is an inverse function of u as expected from basic geometrical optics. A standard blind deconvolution was applied to improve the quality of the images (see Supplementary Information for details).

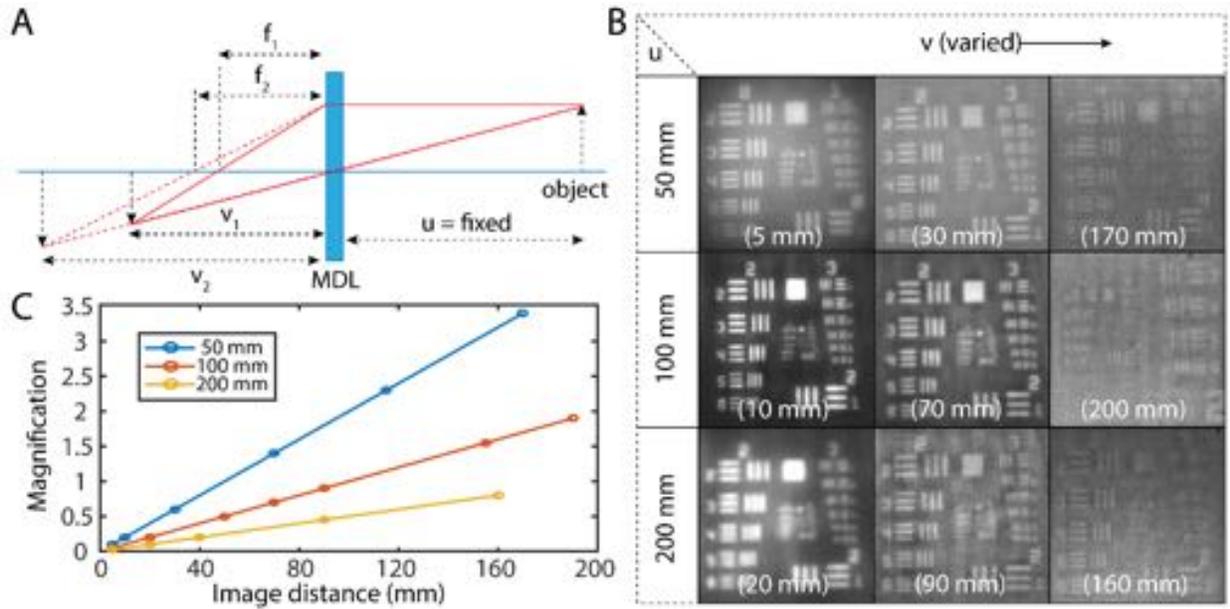

*Figure 4:* *Imaging at different image distances without refocusing. **A.** Focused images of objects at fixed distance u are formed at a large range of image distances, v. **B** Images of the AirForce resolution chart for fixed u and varying v. The value of u is fixed for each row and value of v is noted in parenthesis in each image. **C** Magnification as a function of v for various values of u extracted from the recorded images.*

In the next set of experiments, we kept u fixed and varied v, while recording the images as illustrated in Fig. 4A. These experiments indicate that the image of an object will remain in focus even when the image distance v is changed by a large distance. The recorded images are summarized in Fig. 4B. We note that the MDL is able to form focused images for v from 5mm to 170mm without any change in u. The corresponding magnification was extracted and plotted as a function for v for different values of u in Fig. 4C. Magnification is a linear function of v as expected from basic geometrical optics.

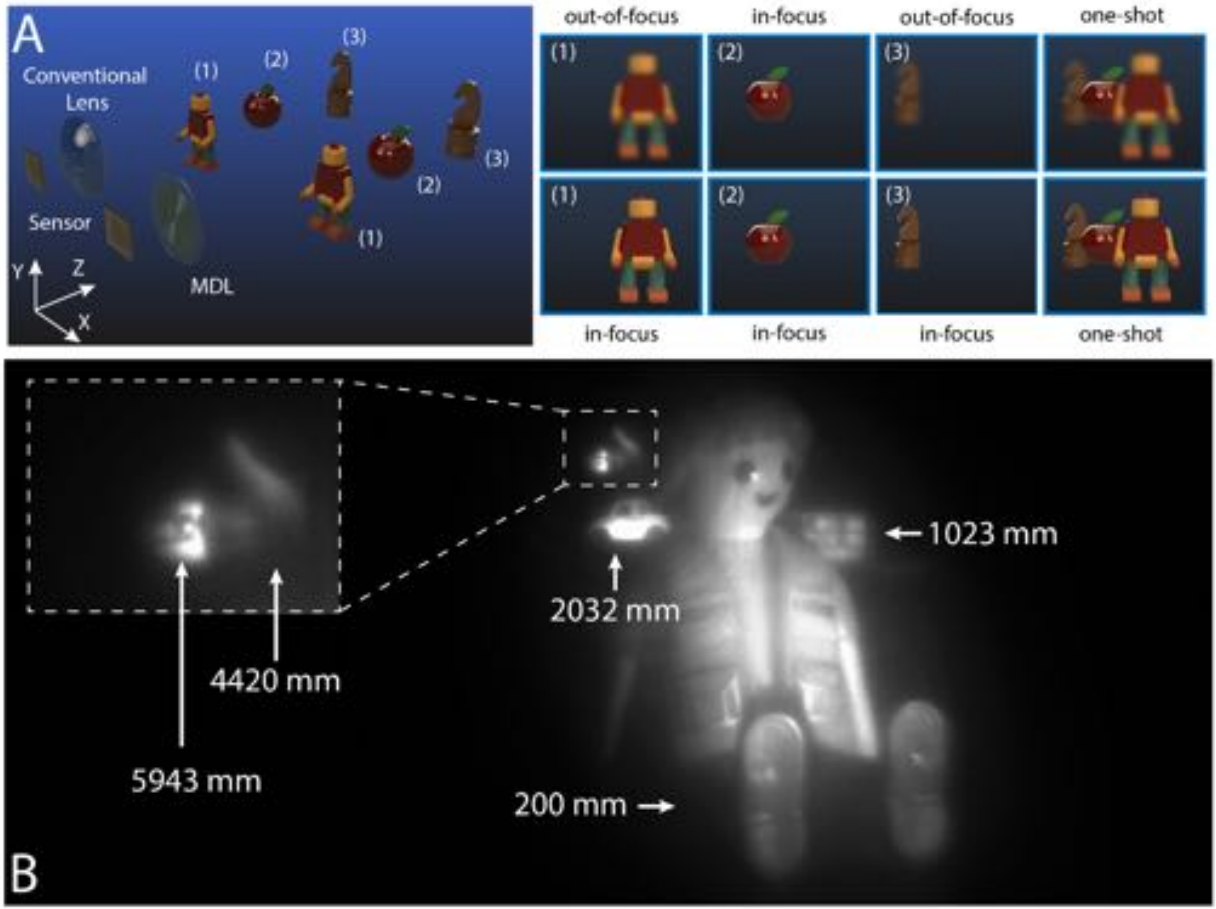

*Figure 5:* *Imaging a scene with large depth of field.* ***A.*** *Objects away from the focus form blurred images with a conventional lens. In contrast, with the MDL, all objects are in focus. Images of the scene taken with the designed MDL. The distance of each object from the MDL (u) is noted in each figure. A video recording of a similar scene is included as supplementary data.*

Finally, in order to demonstrate the potential of our MDL, we imaged a scene containing objects spanning a large depth of field from 200mm to 5943mm. A conventional lens will not be able to keep all the objects in focus over such a large depth of field as illustrated schematically in Fig. 5A. However, the MDL is able to take a single image where all the objects are in focus as shown in Fig. 5B (see section 5 of Supplementary Information for details).

**Conclusion:**

The depth of focus (DOF) is a fundamental property of a lens. By recognizing that the lens is primarily used for intensity imaging, we can treat phase in the image (or focal) plane as a free parameter. Thereby, we can generate a phase-only pupil function that when imprinted on a beam results in a focus that can remain close-to-diffraction limited over a distance that is many orders of magnitude larger than the conventional DOF. Here, we implemented two versions of such pupil functions as multi-level diffractive lenses with a beam focused from 5mm to 1500mm away. Using such lenses we demonstrated extreme DOF imaging of scenes with objects spread as far apart as almost 1000mm apart. By following this design philosophy, we believe that entirely new types of planar optics with large bandwidths, large DOF, etc. become attainable.

**Methods**

**Design and Optimization**

The diffractive lens can be accurately modeled by scalar diffraction theory in the regime of Fresnel approximation. We utilized a modified version of direct-binary search namely gradient descent assisted binary search (GDABS) technique to optimize the surface topography of the MDL.

**Device Fabrication and Characterization**

The lenses were patterned on a photoresist (S1813) film atop a glass wafer using grayscale laser patterning using a Heidelberg Instruments MicroPG101 tool. The exposure dose was varied as a function of position in order to achieve the multiple height levels dictated by the design. After fabrication, the devices were characterized on an optical bench by illuminating them with broadband collimated light, whose spectral bandwidth could be controlled by a tunable filter. The focus of the lenses were captured on a monochrome CMOS sensor for characterization of the PSF. Imaging performance of the lenses were tested in prototype cameras as described in the main text.

## Acknowledgements

We thank Brian Baker, Steve Pritchett and Christian Bach for fabrication advice, and Tom Tiwald (Woollam) for measuring dispersion of materials. We would also like to acknowledge the support from Amazon AWS (051241749381) for help with the computing facilities. RM acknowledges funding from the Office of Naval Research grant N66001-10-1-4065.

## Author Contributions

RM, BSR and SB conceived and designed the experiments. SB and RM modeled and optimized the devices. MM fabricated the device. SB and MM performed the PSF experiments. RM, MM, SB and AM performed the imaging experiments. All authors contributed to the paper.

## Competing Interests Statement

RM is co-founder of Oblate Optics, Inc., which is commercializing technology discussed in this manuscript. The University of Utah has filed for patent protection for technology discussed in this manuscript.

## Materials and Correspondence

Correspondence and materials requests should be addressed to RM at [rmenon@eng.utah.edu](mailto:rmenon@eng.utah.edu).

# Supplementary Information

# Diffractive flat lens enables extreme depth-of-focus imaging

*Sourangsu Banerji, [1] Monjurul Meem, [1] Apratim Majumder, [1] Berardi Sensale-Rodriguez[1] and Rajesh Menon[1, 2, a)]*

[1]*Dept. of Electrical & Computer Engineering, University of Utah, 50 Central Campus Dr. Salt Lake City UT 84112 USA*
[3]*Oblate Optics, Inc. San Diego CA 92130 USA.*

[a)] *rmenon@eng.utah.edu*

# 1. Comparison of Extended Depth of Focus Lenses

| Reference | Operating Wavelength | N.A. | DOF | DOF/Operating wavelength | DOF/diffraction-limited DOF |
|---|---|---|---|---|---|
| 1 | 10.6 um | 0.05 | 40 mm | 3.77E03 | 7.08 |
| 2 | 10.6 um | 0.05 | 37.6 mm | 3.55E03 | 8.86 |
| 3 | 633 nm | 0.0321 | 100 mm | 1.58E05 | 162.7820 |
| 4 | 633 nm | 0.95 | 4.84 um | 7.65 | 3.06 |
| 5 | 633 nm | 0.1724 | 1 mm | ~1.58E03 | 46.9538 |
| 6 | 850 nm | 0.45 | 2.5 mm | ~2.95E03 | 595.5885 |
| 7 | 10.6 um | 0.0504 | 50 mm | ~4.717E03 | 11.98 |
| 8 | 633 nm | 0.93 | 3.16 um | ~5 | 4.32 |
| 9 | 405nm | 0.95 | 4 um | ~10 | 13.53 |
| 10 | 1mm | 0.5 | 248 mm | 248 | 25 |
| 11 | 1mm | 0.6061 | 100mm | 100 | 36.73 |
| 12 | 0.5mm | 0.7 | 10mm | 20 | 9.8 |
| 13 | 1mm | 0.4472 | 100mm | 100 | 19.98 |
| 14 | 3mm | 0.67 | 70mm | ~23.33 | 10.47 |
| 15 | 0.6mm | 0.5812 | 31.3mm | ~52.17 | 17.62 |
| 16 | 1.5 mm | 0.9659 | 48.6mm | ~32.4 | 30.23 |
| 17 | 405nm | 1.2 | 4.05um | 10 | 14.4 |
| 18 | 4.3um | 0.26 | 318um | 74 | 5.02 |
| 19 | N/A | 0.95 | 4 | 4 | 3.61 |
| 20 | 0.6mm | 0.2813 | 120mm | 200 | 15.82 |
| 20 | 3mm | 0.2813 | 120mm | 40 | 9.49 |

## 2. Measured Point Spread Function (PSF)

Some of the measured point spread functions (PSFs) for the designed MDLs are provided below:

**(5mm – 1200 mm):**

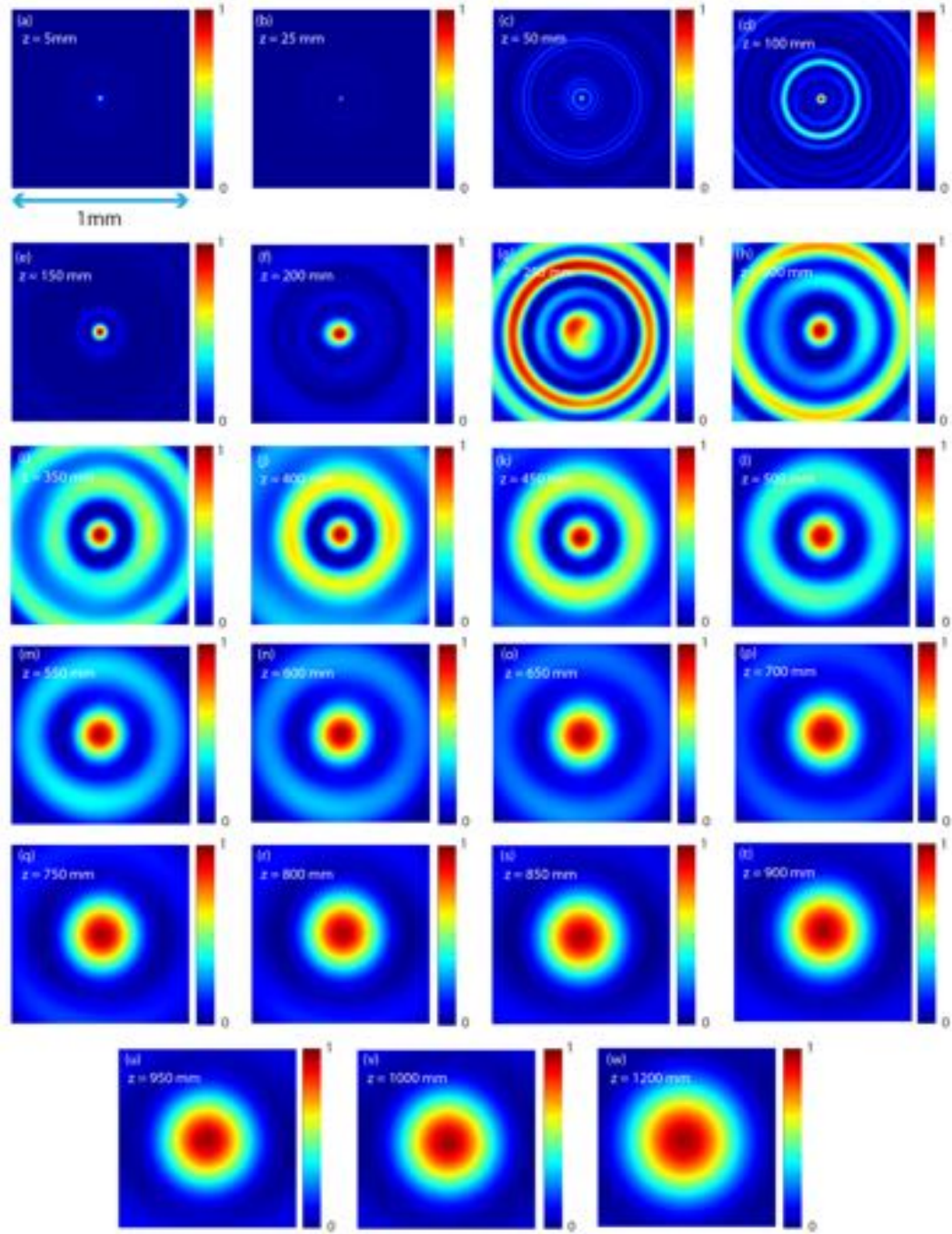

***Fig. S1:*** *Measured PSFs for z =  (a) 5mm (b) 25mm (c) 50mm (d) 100mm (e) 150mm (f) 200mm (g) 250mm (h) 300mm (i) 350mm  (j) 400mm (k) 450mm (l) 500mm (m) 550mm (n) 600mm (o) 650mm (p) 700mm (q) 750mm  (r) 800mm (s) 850 mm  (t) 900mm (u) 950mm (v) 1000mm (w) 1200mm.*

## 3. Efficiency Spectra

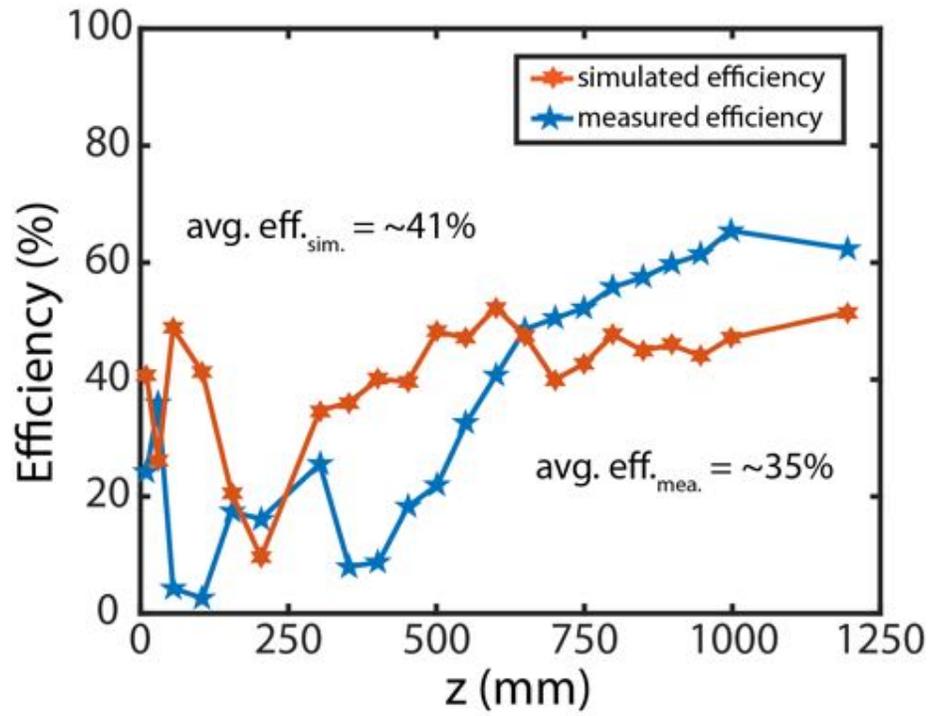

***Fig. S2:*** *Efficiency as a function of the distance z*

## 4. Fabrication

The MDL depicted was patterned in a photoresist (Microchem, S1813) film atop a glass wafer (thickness ~ 0.6 mm) using grayscale laser patterning with a Heidelberg Instruments MicroPG101 tool [21-23]. In such conventional gray-scale lithography, the write head scans through the sample surface and the exposure dose at each point is modulated with different gray-scales [21, 22] (see schematic illustration of Fig. S3). Most of these typical photoresists are characterized by a contrast curve. Different depths in accord with different exposure doses are achieved after development. Greater dose leads to deeper feature. Before patterning structures, it is needed to calibrate this contrast curve. In this case, too, the exposure dose was varied with respect to position to achieve the multiple height levels dictated by the design.

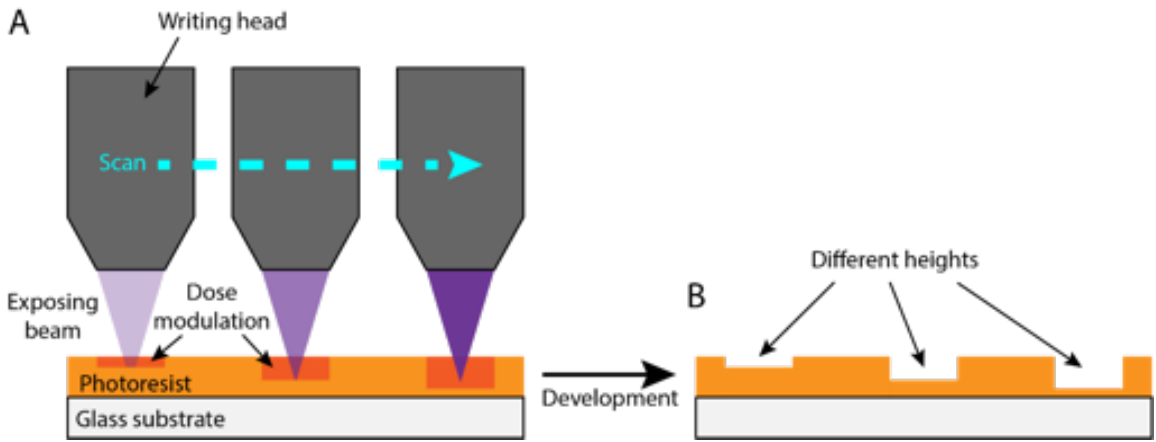

*Fig. S3*: *Schematic illustration of gray-scale lithography on a positive photoresist.*

## 5. Experiment details (focal spot characterization)

The flat lenses were illuminated with expanded and collimated beam from the SuperK VARIA filter (NKT Photonics). The wavelength and bandwidth can be changed using the VARIA filter [24]. The focal planes of the MDLs were magnified using an objective (RMS20X-PF, Thorlabs) and tube lens (ITL200, Thorlabs) and imaged onto monochrome sensor (DMM 27UP031-ML, Imaging Source). The gap between objective and tube lens was ~90 mm and that between the sensor and the backside of tube lens was about 148mm. The magnification of the objective-tube lens was 22.22X.

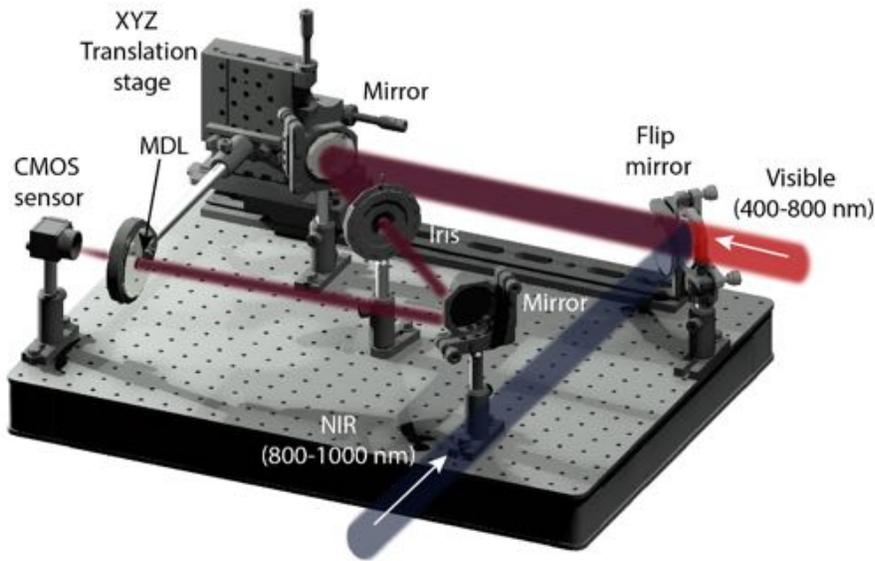

*Fig. S4:* *Schematic of system used for focusing experiments.*

To experimentally determine the focusing efficiency, we used the same setup but now we replaced the monochrome sensor with a 400 μm core diameter fiber tip (P400-1-UV-VIS, Ocean Optics) which in turns was connected to a spectrometer (Jaz Spectrometer, Ocean Optics). The flat lens was illuminated with expanded and collimated beam from the SuperK (850nm). The fiber tip was scanned in X and Y directions using motorized stages so that the fiber tip was aligned with the peak of the magnified focal spot of the flat lens. A slight adjustment in the Z direction was also made to ensure that the integrated signal on the spectroscope was maximum. A reference signal was recorded with light passing through the un-patterned photoresist. Focusing efficiency was then calculated using the following equation: Focusing efficiency = (spectrometer signal when fiber tip aligned to the peak of magnified psf) / [(reference signal) X (area of magnified lens aperture) / (area of fiber tip aperture)]

## 6. Image characterization

The flat lenses were used for imaging the object on to the sensor. The experimental setup is shown in Fig. S5 for both MDLs. The exposure time was adjusted to ensure that the images were not saturated. In each case, a dark frame was recorded and subtracted from the obtained images. For imaging, the objects were placed in front of the MDL. However, this time the objects were illuminated with both IR LEDs and IR floodlights to cover the entire range and the corresponding images were captured using a monochrome sensor (DMM 27UP031-ML, Imaging Source).

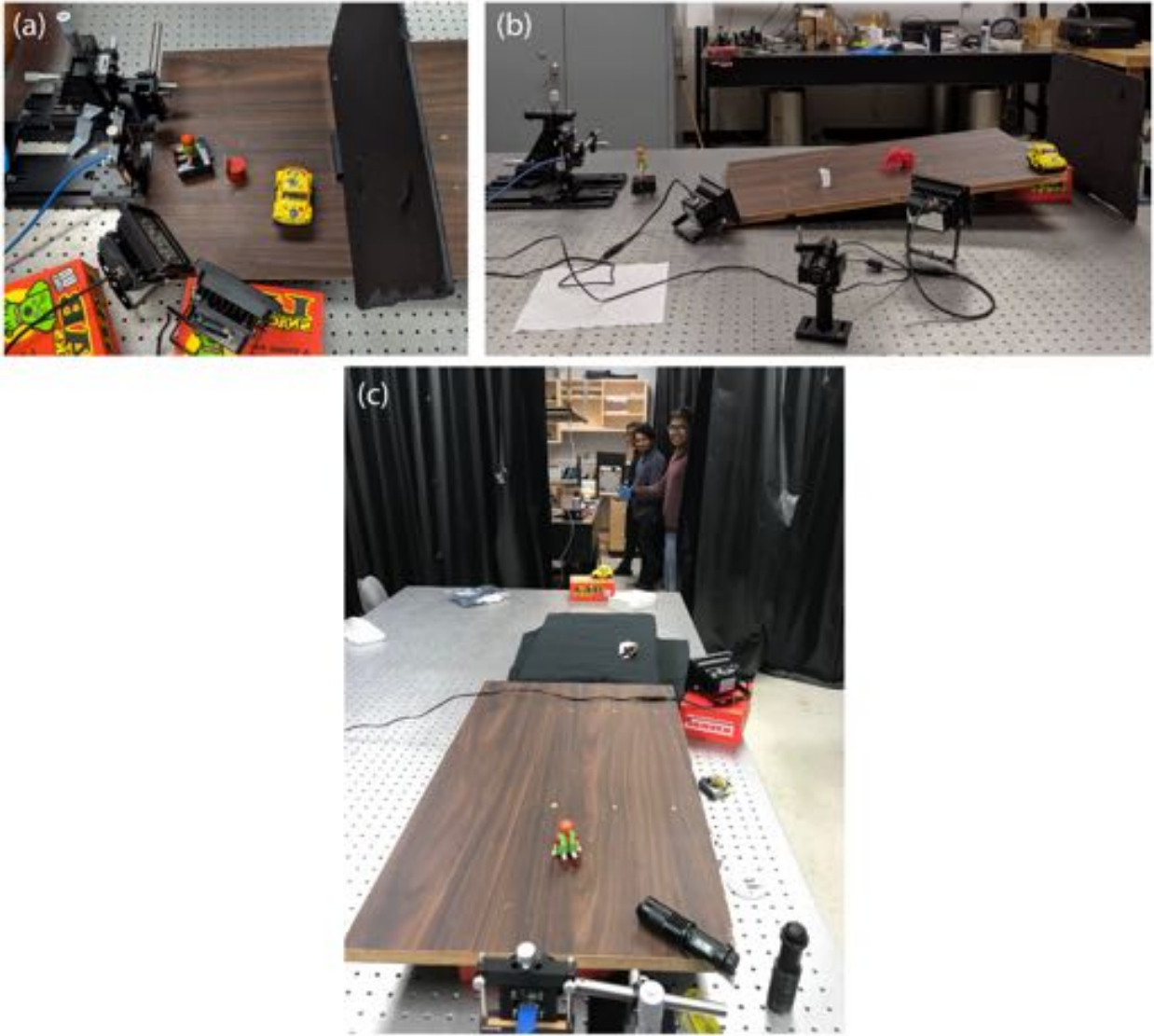

*Fig. S5*: *Experimental setup of image characterization for* **(a)** *MDL1,* **(b)** *MDL2 (1<sup>st</sup> shot) and* **(c)** *MDL2 (2<sup>nd</sup> shot)*

A video showing extreme depth of focus imaging is also included.

**7. Resolution from the USAF 1951 chart**

Resolution test targets are typically used to measure the resolution of an imaging system. They consist of reference line patterns with well-defined thicknesses and spacing, which being designed to be kept in the same plane as the object being imaged. By identifying the largest set of non-distinguishable lines, one determines the resolving power of a given system. The R3L3S1N from Thorlabs (as used here) negative target uses chrome coating to cover the substrate, leaving the pattern itself clear, and works well in back-lit and highly illuminated applications.

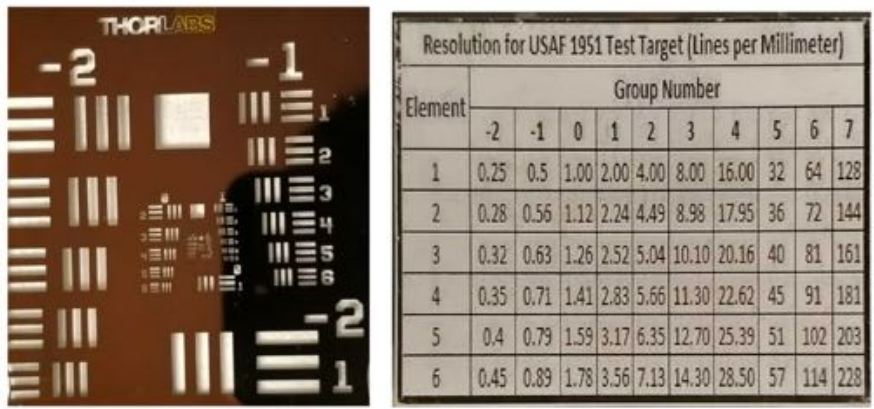

***Fig. S6***: *(a) Resolution Target Chart USAF 1951 (b) Resolution Characterization chart for the resolution target.*

Because these targets feature sets of three lines, they reduce the occurrence of spurious resolution and thus help prevent inaccurate resolution measurements.

## 8. Modulation Transfer Function (MTF) characterization

The average MTF at 10% contrast for the MDL is ~23 lp/mm over the entire range.

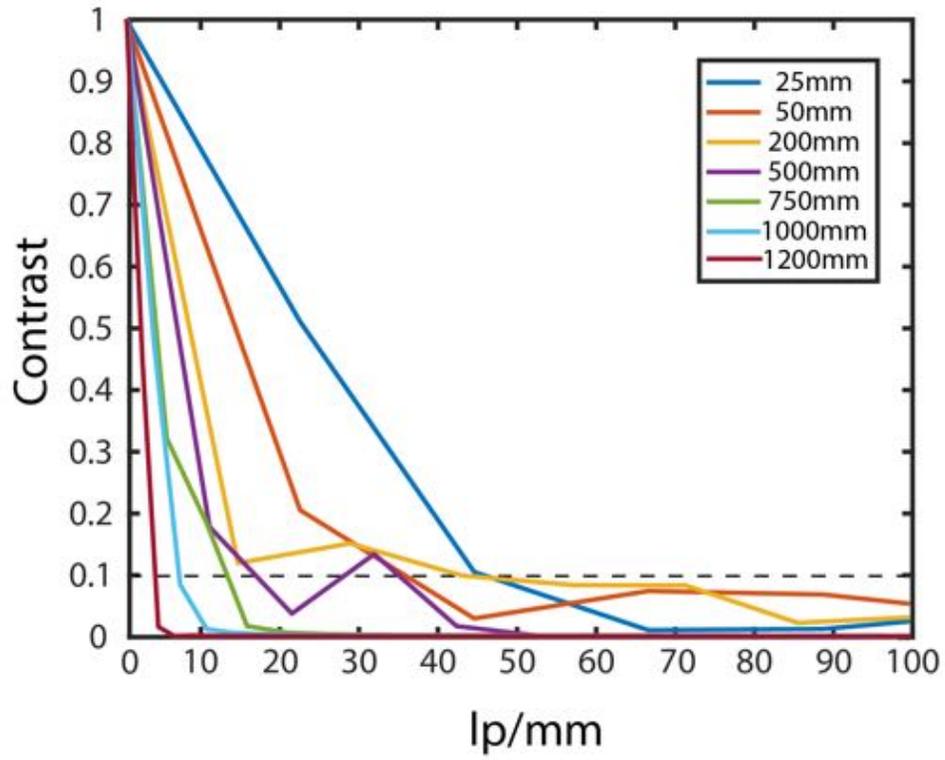

*Fig. S7*: *Modulation Transfer Function for the MDL.*

## 9. Raw Images of the US AirForce resolution chart

**U fixed and V varied**

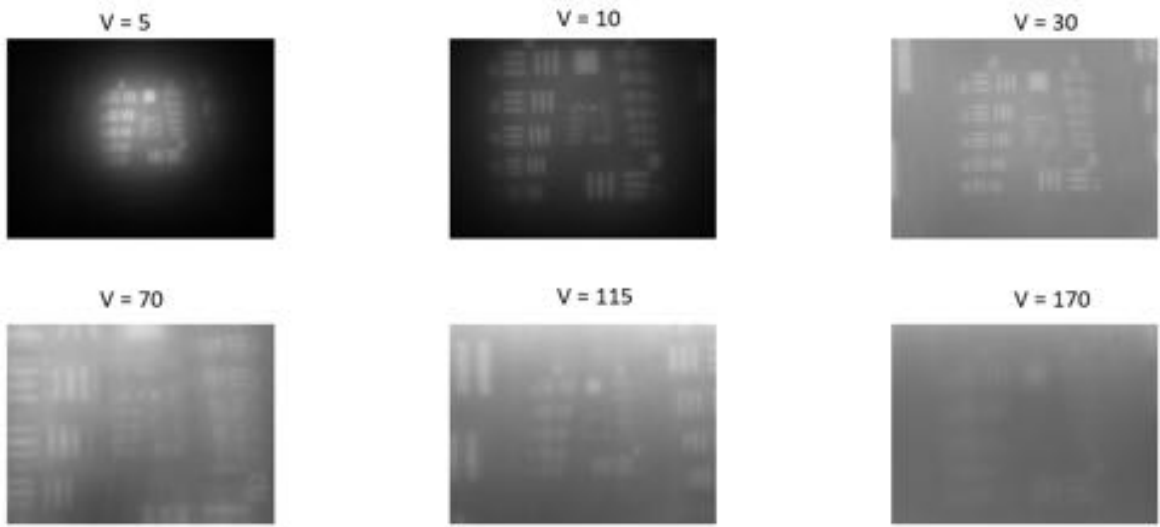
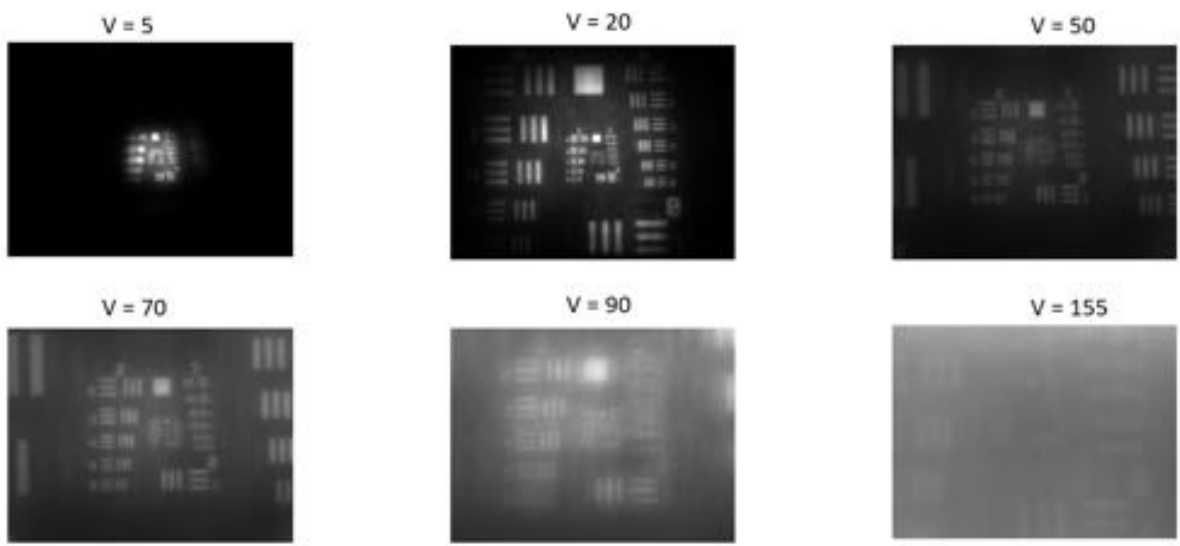

*Fig. 8*: *Resolution Chart images for fixed object distance at 50mm and 100mm and varied image distance*

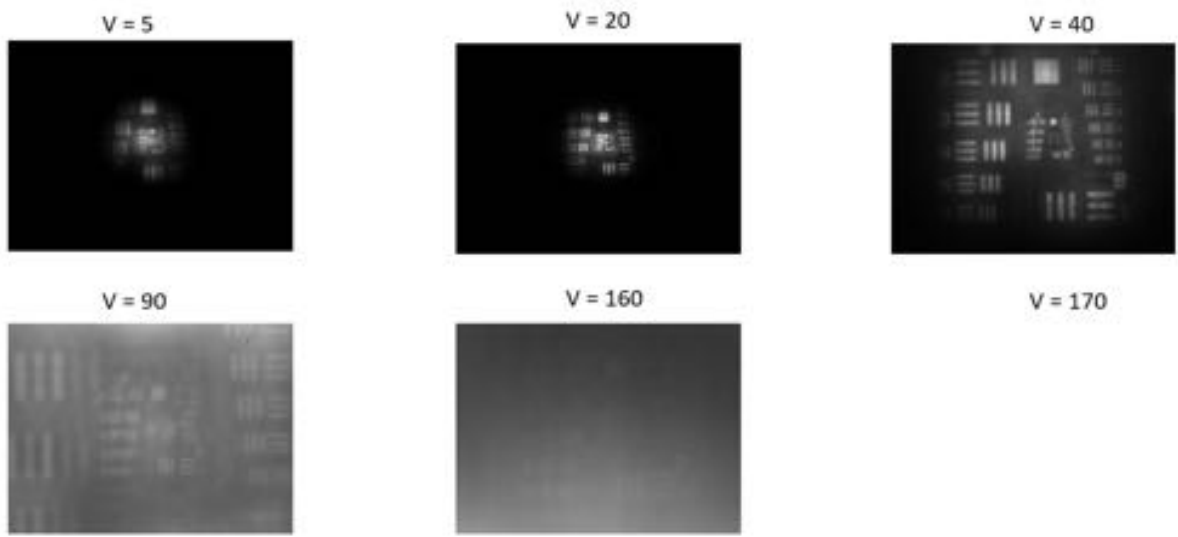

*Fig. 9*: *Resolution Chart images for fixed object distance at 200mm and varied image distance*

**V fixed and U varied**

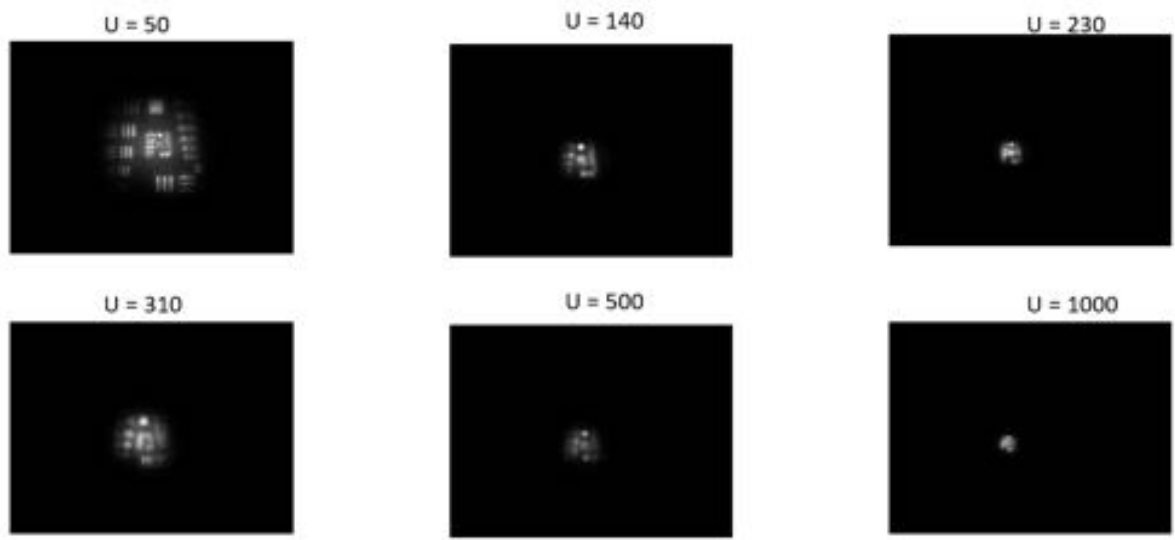

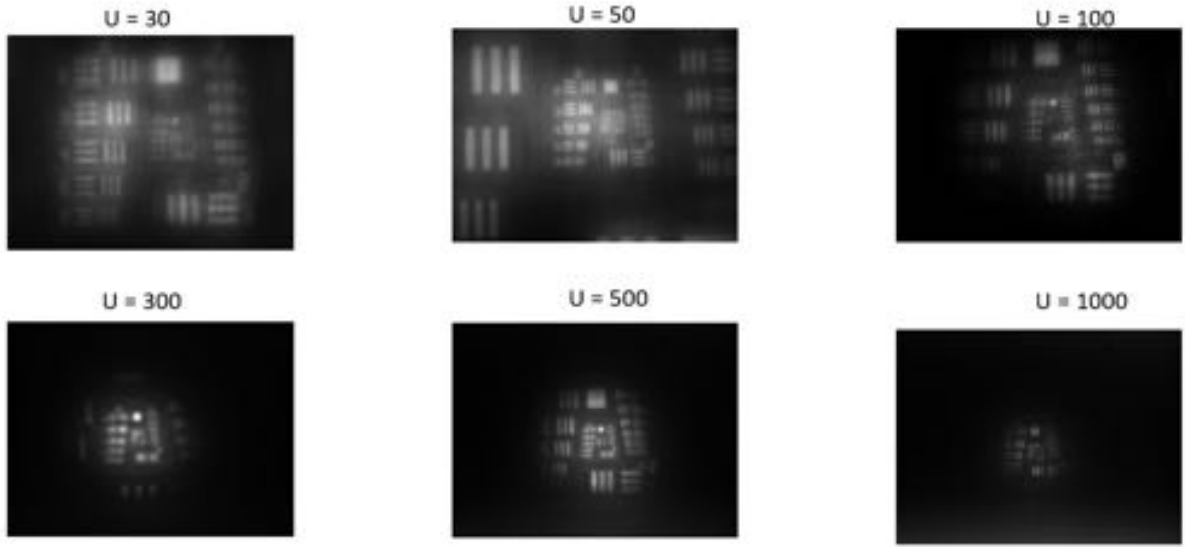

***Fig. 10***: *Resolution Chart images for fixed image distance at 7mm and 20mm and varied object distance*

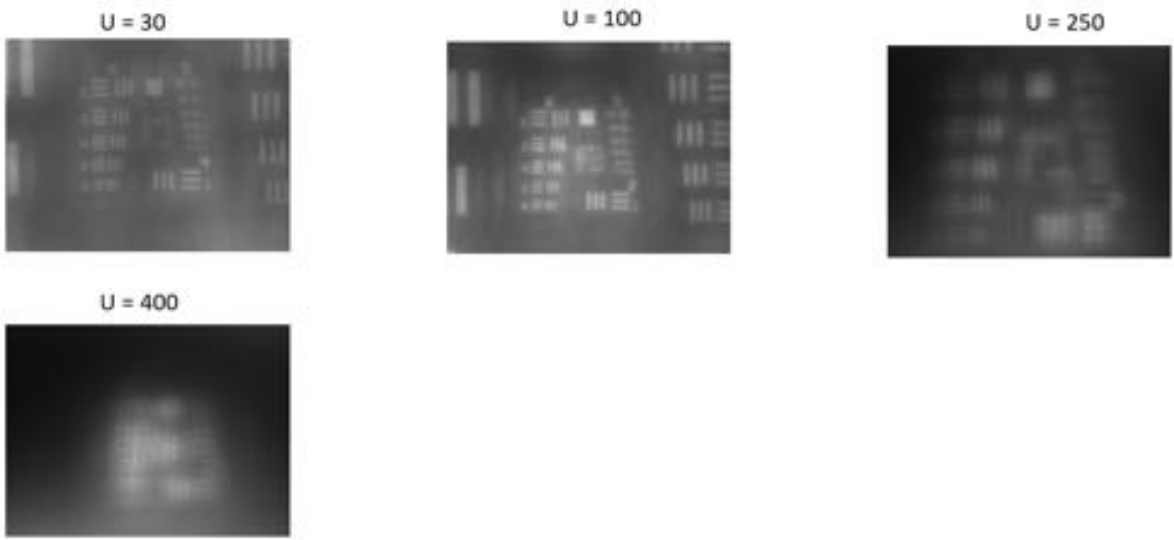

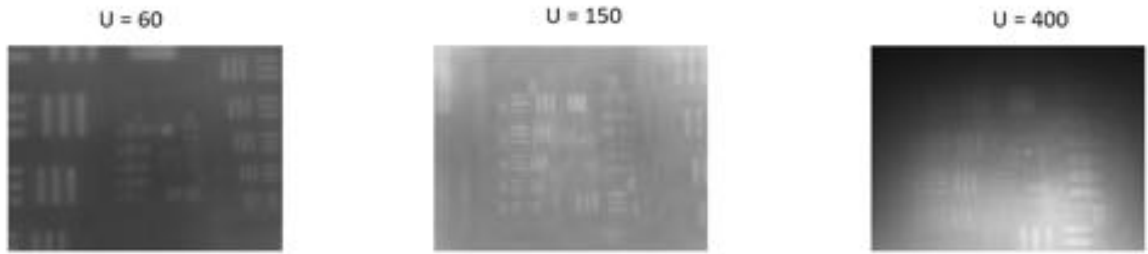

*Fig. 11*: Resolution Chart images for fixed image distance at 50mm and 100mm and varied object distance

## 9. Image taken with a Google Pixel 3 mobile camera

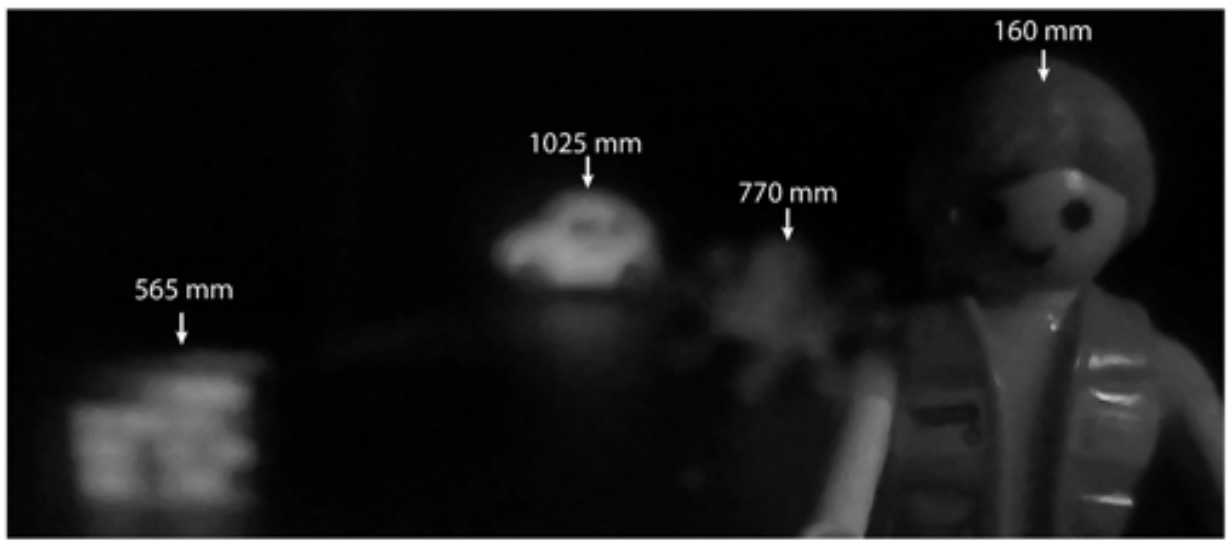

*Fig. 12*: Exemplary image taken with a Google Pixel 3 mobile camera keeping the objects in a relatively similar object distance as show in Fig. 6(c) of the main manuscript.